\documentclass[aps,prd,twocolumn,showpacs,superscriptaddress]{revtex4-1} 
\usepackage[english]{babel}
\usepackage{graphicx,dcolumn,bm,amssymb,amsmath,amsfonts,amsthm,latexsym}
\usepackage{float,slashed,hyperref}
\hypersetup{
  linktocpage  = true,
  colorlinks   = true,
  urlcolor     = black,
  linkcolor    = black,
  citecolor    = blue
}

\newcommand{\be}{\begin{equation}}
\newcommand{\ee}{\end{equation}}
\newcommand{\bea}{\begin{eqnarray}}
\newcommand{\eea}{\end{eqnarray}}
\newcommand{\ba}{\begin{array}{ccc}}
\newcommand{\ea}{\end{array}}
\newcommand{\nn}{\nonumber}

\def\sint{\int \!\!\!\!\!\!\!\!\!\sum_{\ \ np}\ }

\begin{document}

\title{Time-dependent Ginzburg-Landau approach to the dynamics of inhomogeneous chiral condensates in a nonlocal Nambu--Jona-Lasinio model}

\author{J.P.~Carlomagno}
\email{carlomagno@fisica.unlp.edu.ar}
\affiliation{IFLP, CONICET $-$ Departamento de F\'{i}sica, Universidad Nacional de La Plata, C.C. 67, 1900 La Plata, Argentina}
\affiliation{CONICET, Rivadavia 1917, 1033 Buenos Aires, Argentina}
\affiliation{ICTP South American Institute for Fundamental Research, Rua Dr.~Bento Teobaldo Ferraz, 271 - Bloco II, 01140-070 S\~ao Paulo, SP, Brazil}
\author{Gast\~{a}o Krein}
\email{gkrein@ift.unesp.br}
\affiliation{Instituto de F\'isica Te\'orica, Universidade Estadual Paulista, \\ Rua Dr.~Bento Teobaldo Ferraz, 271 - Bloco II, 01140-070 S\~ao Paulo, SP, Brazil} 


\begin{abstract}
We study the dynamics of inhomogeneous scalar and pseudoscalar chiral order parameters within the framework of the time-dependent Ginzburg-Landau equations. 
We utilize a nonlocal chiral quark model to obtain the phase diagram of the model as function of temperature and baryon chemical potential and study the formation of metastable spatial domains of matter where the order parameters acquire a spatial modulation in the course their dynamical evolution. 
We found that, before reaching the expected equilibrium homogeneous state, both scalar and pseudoscalar chiral condensates go through long-lived metastable inhomogeneous structures. 
For different initial configurations of the order parameters, the lifetimes of the inhomogeneous structures are compared to timescales in a relativistic heavy-ion collision.
\end{abstract}

\maketitle

%
\section{Introduction}
\label{intro}

The phase diagram of strongly interacting matter at finite temperature ($T$) and baryon chemical potential ($\mu$) has been extensively studied along the last decades. 
Quantum Chromodynamics (QCD) predicts that at very high temperatures ($T \gg \Lambda_{\rm QCD}$) and low baryon densities this matter appears in the form of a plasma of quarks and gluons, and at very high baryon densities ($\mu \gg \Lambda_{\rm QCD}$) and zero temperatures it is a color superconductor~\cite{Fukushima:2010bq}. 
At such extremes, QCD is weakly coupled and first-principles perturbative calculations based on an expansion in the coupling constant can be used to explore the phase diagram. 
But there are regions of the phase diagram with temperatures and densities interpolating the extremes that remain poorly understood. 
An example is the region of low temperatures and densities close to or a few times larger than the nuclear saturation density, which is of great interest for the physics of compact stars~\cite{Holt:2014hma,Baym:2017whm,Blaschke:2018mqw}. 
Here, QCD is strongly coupled and coupling-constant expansions become inapplicable. 
First-principles nonperturbative lattice QCD methods based on large-scale Monte Carlo simulations are also not applicable in this case, because at finite $\mu$ QCD has a {\em sign problem}~\cite{Splittorff:2007ck,Aarts:2015tyj}. 

Lattice QCD has established, in particular, the existence of a finite-temperature crossover for chiral symmetry restoration at vanishing small densities~\cite{Aoki:2006we,Cheng:2007jq, Bhattacharya:2014ara}. 
In vacuum, the approximate chiral symmetry of the QCD Lagrangian in the light quark sector is dynamically broken, a feature that explains the lightness of the pion and is responsible for generating the bulk of the masses of the light hadrons, like protons and neutrons~\cite{Nambu:1960xd,Horn:2016rip}. 
On the other hand, the behavior of chiral symmetry when moving from vacuum to densities of ordinary nuclear matter and higher is not well understood and all of what is presently known comes from model calculations. 
In this context, recent works~\cite{Nakano:2004cd,Basar:2008im,Basar:2008ki,Nickel:2009ke,Basar:2009fg,Nickel:2009wj} have revived the discussion on the possibility that chiral symmetry breaking in dense matter at low temperatures would drive the formation of nonuniform phases, i.e. the formation of spatially-varying chiral condensates which break translational 
invariance{\textemdash}Ref.~\cite{Buballa:2014tba} is a thorough recent review on the subject, with an account on earlier developments and a large list of references. 
One particularly interesting result suggests that, in addition to the expected tricritical endpoint of the first order chiral phase transition, there might exist a Lifshitz point, where two homogeneous phases and one inhomogeneous phase meet~\cite{Nickel:2009ke,Nickel:2009wj,Carignano:2014jla}. 
Interestingly, also for ordinary cold nuclear matter there seems to be the  possibility that an inhomogeneous chiral phase appears at densities a few times larger than the normal density~\cite{Heinz:2013hza}. 
These are interesting features of dynamical chiral symmetry breaking in matter and it would be fascinating to find signals of their existence in real systems. 

Recent studies~\cite{Tatsumi:2014cea,Carignano:2015kda,Buballa:2015awa} have suggested possible observable signals of the presence of inhomogeneous phases in hybrid stars.
In Ref.~\cite{Tatsumi:2014cea} a novel cooling scenario was suggested, in that inhomogeneities induce modifications in momentum-conservation relations in quark beta decay, leading to neutrino emissivities with efficiencies comparable to those due to interacting quarks or due to the presence of a pion condensate. 
Refs.~\cite{Carignano:2015kda,Buballa:2015awa} investigated the consequences of an inhomogeneous chiral phase for the equation of state of matter in a hybrid star, finding substantial effects on the mass-radius relation of the star. 
Clues on such phases are also expected from experiments of heavy-ion collisions, like those ongoing within the beam energy scan program at Relativistic Heavy Ion Collider (RHIC)~\cite{Kumar:2013cqa} and also from those planned to be conducted at the Facility for Antiproton and Ion Research (FAIR)~\cite{Senger:2017nvf}, the Nuclotron-based Ion Collider Facility (NICA)~\cite{Senger:2016wfb} and the Japan Proton Accelerator Research Complex (J-PARC)~\cite{Sako:2017chd}. 
If metastable inhomogeneous phases are present in the matter produced in a heavy-ion collision, the system would spend different   times in the different phases during its evolution and therefore, this information could in principle be recovered employing for instance the freeze-out eccentricity which provides  geometric information~\cite{Kumar:2013cqa}. 
There has also been the suggestion that (thermal and quantum) fluctuations around a mean-field inhomogeneous condensate induce anomalies in thermodynamic quantities that could be revealed in particle production yields~\cite{Yoshiike:2017kbx}. 

Parallel to the quest of observable signals in heavy-ion collisions, there is the important issue regarding the dynamical evolution of an inhomogeneous chiral configuration in this context.
Particularly important is the time scale associated with the nonequilibrium evolution of such a configuration from its formation until its decay into freely steaming particles (mostly pions) that eventually will reach the detectors. 
Clearly, a first-principles theoretical study of this issue is currently out of reach. 
Nevertheless, likewise with the equilibrium situation, useful qualitative insight on the nonequilibrium evolution can be obtained through the Ginzburg-Landau (GL) mean-field approach. 
In this approach, given the GL functional, i.e. the thermodynamic potential, functional of the order parameters, the phase structure of the system is obtained by exploring the extrema of the potential with respect to the order parameters.
When the system is brought out of equilibrium, the relaxation of the order parameters, i.e. their time evolution towards to an equilibrium state is described by the time-dependent Ginzburg-Landau (TDGL) equation~\cite{Onuki:2002}{\textemdash}a good review on the TDGL approach for condensed matter systems can be found in Ref.~\cite{Bray:1994zz}. This approach and variants of it have been extensively used along the last decades to investigate different aspects of the relaxation dynamics of chiral order parameters and also order parameters associated with conserved charges and color confinement~\cite{Csernai:1995zn,Biro:1997va,Mishustin:1998yc,Scavenius:1999zc,Miller:2000pd,Bower:2001fq,Scavenius:2001pa,Paech:2003fe,Aziz:2004qu,Fraga:2004hp,Paech:2005cx,Fraga:2006cr,Koide:2006vf,Fraga:2007gg,CassolSeewald:2007ru,Randrup:2010ax,Singh:2011rz,Kapusta:2012zb,Singh:2012nq,Singh:2013pxa}. 
In the present paper, to get insight into the dynamics of inhomogeneous chiral condensates, we follow those lines and employ the TDGL approach. 
In general, to obtain the $T-$ and $\mu-$dependence of the GL functional, a model is required. Here we employ a nonlocal Nambu$-$Jona-Lasinio model (nlNJL)~\cite{Carlomagno:2014hoa,Carlomagno:2015nsa}, in which quark fields interact through a non local chiral invariant four-fermion coupling, to obtain the thermodynamic energy functional. 
We focus on the time evolution of the the chiral order parameters at low values of the $T$, and different values of $\mu$, both close to the tricritical and the Lifshitz points, as predicted by that model.

The paper is organized as follows. 
In the next section we define the nonlocal NJL model we use and briefly review the TDGL approach. Section~\ref{results} presents the results of the numerical simulations of the TDGL equations. Finally, Conclusions and Perspectives are presented in section~\ref{finale}.

%
\section{TDGL approach}
\label{nlnjl}

As in Ref.~\cite{Carlomagno:2014hoa}, we consider the simplest version of a nonlocal SU(2) NJL model in the chiral limit. 
The corresponding Euclidean effective action is given by
\be
S_{E}= \int d^{4}x\
\left[ -i \bar{\psi}(x) \, \rlap/\partial \, \psi(x)
-\frac{G}{2} \ j_{a}(x) \, j_{a}(x)  \right] ,
\label{action}
\ee
where $\psi$ stands for the $N_{f}=2$ fermion doublet $\psi = (u,d)^T$, and 
$j_{a}(x)$ for the nonlocal currents
\bea
j_{a}(x)  =\int d^{4}z\, {\cal G}(z)\, \bar{\psi}\left(x+\frac{z}{2}\right)
\, \Gamma_{a}\, \psi\left(  x-\frac{z}{2}\right) , 
\label{currents}
\eea
where we have defined $\Gamma_{a}=(\leavevmode\hbox{\small1\kern-3.8pt\normalsize1}, i\gamma_{5}\vec{\tau})$, and the function ${\cal G}(z)$ is a nonlocal form factor that characterizes the effective interaction.

To proceed, we perform a standard bosonization of the theory, in which bosonic fields are introduced and quark fields are integrated out. 
Within the GL approach, the bosonic fields are replaced by their (vacuum or thermodynamic) expectation (or mean-field) values $\phi(\vec x)=(\sigma (\vec x), \vec \pi (\vec x))$. 
Since now parity is not necessarily an exact symmetry, one can get in general a nonzero value for the pseudoscalar field. 
The mean-field values are allowed to be inhomogeneous, hence the explicit dependence on spatial coordinates. 
The phase structure of the system is obtained by exploring the extrema of the GL functional, the thermodynamic potential $\Omega_{\rm GL}$ obtained from the action in Eq.~(\ref{action}), functional of the order parameter $\phi$:
\be
\dfrac{\delta \Omega_{GL}[\phi]}
{\delta \phi (\vec x)} = 0.
\label{mf-stat}
\ee
This gives a coupled system of equations for $\sigma$ and $\pi$, whose solutions reveal the presence, or absence, of inhomogeneous configurations of the fields $\sigma$ and $\pi$ in thermodynamic equilibrium at given values of $T$ and $\mu$. 
The question to be addressed is the time evolution of an initial configuration $\phi$, which is not a fully-developed equilibrium configuration. 
Within the TDGL framework, the time evolution of $\phi$ is governed by the dynamical equation~\cite{Onuki:2002}:
\be
\Gamma\ \dfrac{\partial \phi (\vec x,t)}{\partial t} = -  \dfrac{\delta \Omega_{GL}[\phi]}
{\delta \phi (\vec x,t)} ,
\label{tdgle}
\ee
where $\Gamma$ is a kinetic coefficient due to dissipation processes, like $\sigma \leftrightarrow 2 \pi$ and $\pi \sigma \rightarrow \pi$; it is $T-$ and $\mu-$dependent and, in principle, different for $\sigma$ and $\pi$. 
The equilibrium configurations are the $\partial \phi/\partial t =0$ stationary solutions for $t \rightarrow \infty$, i.e. the mean-field equation Eq.~(\ref{mf-stat}). 
The equilibrium configurations are independent of $\Gamma$, but the time scale for them to become established is governed by $\Gamma$.

We note that thermal fluctuations are not taken into account in Eq.~(\ref{tdgle}). 
Although they are small for low temperatures, as in the case we are going to study in the present paper, they play an important role at temperatures e.g. close to the crossover temperature.  
Fluctuations are usually taken into account phenomenologically by adding noise terms on the right hand side of Eq.~(\ref{tdgle}), turning the TDGL equation into a stochastic equation~\cite{Biro:1997va,Miller:2000pd,CassolSeewald:2007ru,Randrup:2010ax,Kapusta:2012zb}. 
The strength of the noise fields are constrained by the dissipation-fluctuation theorem.
Equations of this sort can be derived from a microscopic model via influence functional techniques or the closed-time-path (CTP) effective action formalism~\cite{Calzetta:2008iqa}; in general, they contain nonlocal dissipation kernels and colored noise fields{\textemdash}see e.g. Refs.~\cite{Gleiser:1993ea,Boyanovsky:1994me,Greiner:1996dx,Greiner:1998vd,Rischke:1998qy,Nahrgang:2011mv,Nahrgang:2011mg,Gautier:2012vh}. 
We also note that Eq.~(\ref{tdgle}) is a TDGL equation for nonconserved order parameters, like for $\sigma$ and $\pi$ in the present case; for conserved order parameters, like the baryon density, different phenomenological equations are used~\cite{Onuki:2002}. 

As mentioned earlier, the situation of interest is the time evolution of the order parameters starting from initial configurations for which chiral symmetry is not fully restored. 
The physical picture behind such a scenario resembles the cooling stage in the course of the evolution of a heavy-ion collision toward a state of broken chiral symmetry.
To make contact with the studies at equilibrium in the literature, we follow Refs.~\cite{Nickel:2009ke,Carlomagno:2014hoa} and expand the mean-field thermodynamic potential in powers of the order parameters and their spatial gradients as:
\be
\Omega_{\rm GL}[\phi] = \int d^3x \, \omega_{\rm GL}(T,\mu,\phi),
\ee
where $\omega_{\rm GL}(T,\mu,\phi)$ is the GL energy-density functional
\begin{widetext}
\bea
\omega_{\rm GL}(T,\mu,\phi) & = & \frac{\alpha_2}{2} \ \phi^2
+ \frac{\alpha_4}{4} (\phi^2)^2 + \frac{\alpha_{4b}}{4} (\nabla \phi)^2 
+ \frac{\alpha_6}{6} (\phi^2)^3 + \frac{\alpha_{6b}}{6} (\phi, \nabla \phi)^2 
+ \, \frac{\alpha_{6c}}{6} \left[ \phi^2 (\nabla \phi)^2 - (\phi, \nabla \phi)^2 \right] ,
\label{genome}  
\eea
with the expansion coefficients $\alpha_2, \cdots,\alpha_{6d}$, which are functions of $T$ and $\mu$, given by
\bea
\alpha_2 &=& \frac{1}{G} - 8 \, N_c \, \sint \frac{g^2}{p_n^2},
%
\hspace{1.0cm}
\alpha_4 = 8 \, N_c \, \sint \frac{g^4}{p_n^4}, \hspace{1.0cm}
\alpha_6 = - 8 \, N_c \, \sint \frac{g^6}{p_n^6},
\nn \\
\alpha_{4b} &=& 8 \, N_c \, \sint \frac{g^2}{p_n^4} \left(1 -
\frac{2}{3}\,\frac{g'}{g}\, \vec p^{\;2}\right), 
%
\hspace{1.0cm}
\alpha_{6b} = -40 \, N_c \, \sint \frac{g^4}{p_n^6} \left( 1 -
\frac{26}{15}\,\frac{g'}{g} \, \vec p^{\; 2}
+ \frac{8}{5} \, \frac{{g'}^2}{g^2}\, \vec p^{\; 2} p_n^2 \right),
\nn \\
\alpha_{6c} &=& -24 \, N_c \, \sint \frac{g^4}{p_n^6} \left( 1 - \frac{2}{3}\,\frac{g'}{g}\,
\vec p^{\; 2} \right),
%
\hspace{1.0cm}
\alpha_{6d} = -4 \, N_c \, \sint \frac{g^2}{p_n^6}
\left[ 1 - \frac{2}{3}\,\frac{g'}{g}\,\vec p^{\; 2} + \frac{1}{5}
\left(\frac{{g'}^2}{g^2}+\frac{g''}{g}
\right)\vec p^{\; 4}\right],
\label{glcoef}
\eea
where we have used the shorthand notation
\be
\sint  \equiv \ \frac{T}{2 \pi^2}\sum_{n=-\infty}^{\infty}
\int_0^\infty d|\vec p|\; \vec p^{\; 2}, \label{nonloc}
\ee
and $p_n^2 \equiv \left[ (2n+1) \pi T - i \, \mu \right]^2 + \vec p^{\; 2}$. 
The function $g$, evaluated at $p^2 = p_n^2$, is the Fourier transform of the non local form factor ${\cal G}(x)$ of the quark-antiquark currents, and $g'$ and $g''$ denote derivatives with respect to $\vec p^{\; 2}$. 
We recall that the GL expansion in powers of the order parameter and its gradients, is expected to be valid only close to the second order transition to the chirally restored phase~\cite{Fulde:1964zz,Alford:2000ze,Abuki:2011pf,Carignano:2017meb}. 
Moreover, at the Lifshitz point, both the order parameter and its gradient are expected to vanish. 
As in Refs.~\cite{Nickel:2009ke,Carlomagno:2014hoa}, we also restrict the analysis to one-dimensional modulations of the order parameters. 
In this case, the two TDGL equations for $\sigma$ and $\pi$ are given by
\bea
\Gamma_\sigma \, \dfrac{\partial \sigma (\vec x,t)}{\partial t}  &=& 
- \alpha_2 \sigma - \alpha_4 (\sigma^2 + \pi^2) - \alpha_6 (\sigma^2 + \pi^2)^2 \sigma 
+ \alpha_{4b} \sigma'' +  \frac{\alpha_{6b}}{3} \left(\sigma'^2 + \pi'^2 + \sigma \sigma' 
- \pi \pi' \right) \sigma  \nn \\
&& - \, \frac{\alpha_{6c}}{3} \left[(2 \pi'^2 + \pi \pi'') \sigma - 2 \pi \pi' \sigma' 
+ \pi^2 \sigma'' \right] - \frac{\alpha_{6d}}{3} \sigma^{(4)} ,
\label{gllang1} \\
\Gamma_\pi \, \dfrac{\partial \pi (\vec x,t)}{\partial t}  &=&  - \alpha_2 \pi 
+ \alpha_4 (\pi^2 + \sigma^2) - \alpha_6 (\pi^2 + \sigma^2)^2 \pi 
+ \alpha_{4b} \pi'' 
+ \frac{\alpha_{6b}}{3} \left(\pi'^2 + \sigma'^2 + \pi \pi' + \sigma \sigma' \right) \pi 
\nn \\
&& - \, \frac{\alpha_{6c}}{3} \left[(2 \sigma'^2 + \sigma \sigma'') \pi + 2 \sigma \sigma' \pi' 
+ \sigma^2 \pi'' \right] - \frac{\alpha_{6d}}{3} \pi^{(4)},
\label{gllang2}
\eea
where the primes denote spatial derivatives. 
\pagebreak
\end{widetext}

To conclude this section, we note that for a local, contact-interaction model, ${\cal G}(x) \sim \delta(x)$, the $\alpha_n$ are reduced to
\bea
\alpha_{4b} = \alpha_4, \hspace{0.5cm}
\alpha_{6b}/5 = \alpha_{6c}/3 = 2\, \alpha_{6d} = \alpha_6 . 
\label{njl}
\eea

%
\section{Numerical Results}
\label{results}

In the chiral limit, the model is completely determined by the form factor $g(p)$ and the coupling constant $G$. 
We choose the Gaussian form for $g(p)$, considered in many previous studies~\cite{Bowler:1994ir,Schmidt:1994di,Golli:1998rf,GomezDumm:2001fz,Scarpettini:2003fj,GomezDumm:2004sr,GomezDumm:2006vz}:
\be
g(p) = \exp(- p^2/\Lambda^2),
\label{gaussian}
\ee
where $\Lambda$ is the range of the interaction in momentum space. 
It is usual to fix those parameters so as to get phenomenologically adequate values for the pion decay constant and the quark-antiquark condensate. 
Here, we use $f_\pi = 86$~MeV and $\langle\bar q q\rangle = -(270$~MeV$)^3$~\cite{Aoki:2013ldr}, to determine $\Lambda$ and $G$.
They are given by $G= 14.668$~GeV$^{-2}$ and $\Lambda = 1.046$~GeV. 

\begin{figure}[H]
\centering
\includegraphics[width=0.83\columnwidth]{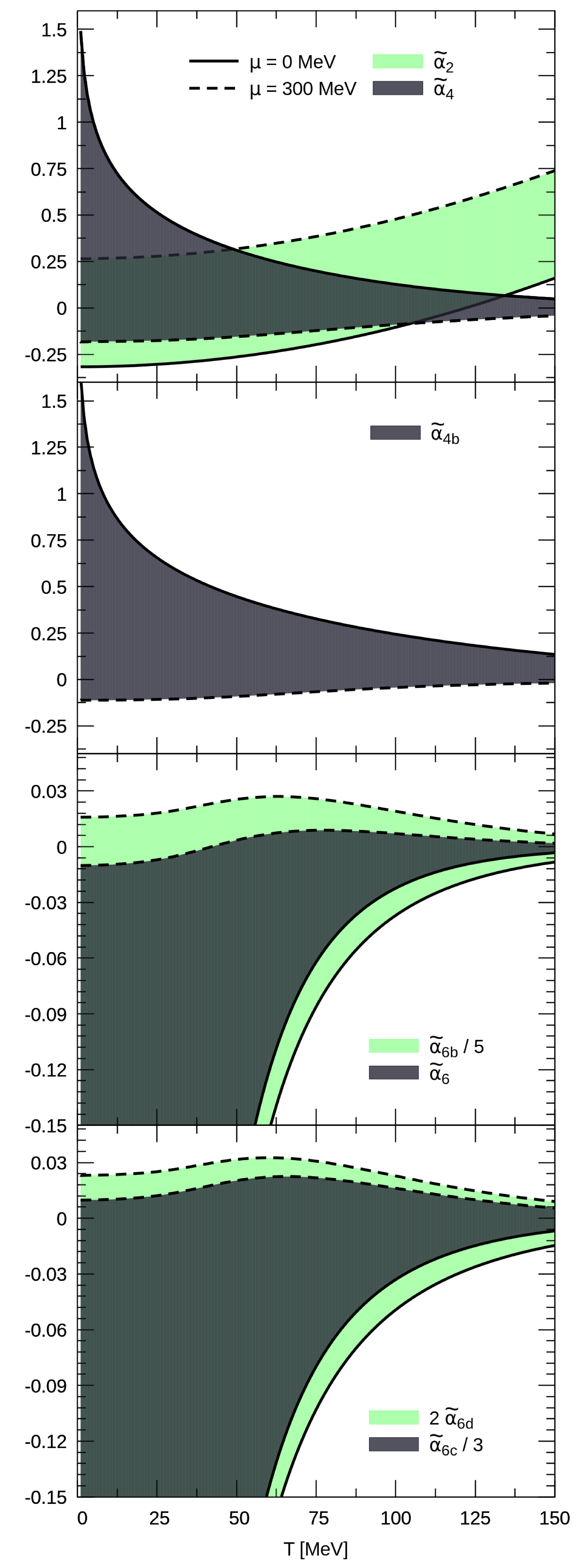}
\caption{\label{fig:alfa} Normalized GL coefficients as a function of temperature, for values of $\mu$ varying between $\mu=0$ and $300$~MeV, enclosed by the solid and dashed lines, respectively.}
\end{figure}

The explicit $T$ and $\mu$ dependence of the GL coefficients $\alpha_2, \cdots,\alpha_{6d}$ have not been presented previously for a nonlocal model, therefore in Fig.~\ref{fig:alfa} we plot them as a function of temperature for different values of the chemical potential. 
The figure shows the dimensionless coefficients $\widetilde{\alpha}_n$, which are ratios of ${\alpha}_n$ to the appropriate powers of the vacuum quark condensate $\langle\bar q q\rangle$.  
Solid and dashed lines, enclosing the shaded areas, correspond to $\mu=0$ and $300$~MeV, respectively. 

Except for $\widetilde{\alpha}_2$ (upper panel), we see that the larger the temperature, the smaller are the magnitudes of the $\widetilde{\alpha}_n$. 
Furthermore, the sign of the coefficients is strongly $\mu$ dependent. 
The energy density as a function of the order parameter for given values of $T$ and $\mu$ is defined by the magnitude of the GL coefficients.
Moreover, the analysis of the relative sign between those coefficients determines the different regions in the QCD phase diagram. 
In particular, the Lifshitz point (LP) is defined as the point where the inhomogeneous phase and the two homogeneous phases with broken and restored chiral symmetry meet, while the tricritical point (TCP) denotes the point where the second-order chiral phase transition turns into a first order one. 
The positions of the LP and the TCP can be determined by solving, as a function of $T$ and $\mu$, the following set of equations~\cite{Buballa:2014tba, Carlomagno:2014hoa}
\bea
\alpha_2 = 0 , \qquad \alpha_{4b} = 0 , \qquad \alpha_2 = 0 \ , \qquad \alpha_4 = 0 .
\label{eqlptcp}
\eea
For the set of parameters of the nonlocal NJL model given above, the coordinates in the $(T,\mu)$ plane for the LP and the TCP (in~MeV) are $(35.78,211.63)$ and $(74.83,174.86)$, respectively.
The reader is referred to Ref.~\cite{Carlomagno:2015nsa}, where, for a particular spatial modulation, phase diagrams for different set parameters (fixed by different values of the quark condensate) are shown. 
In particular, for $\langle\bar q q\rangle = -(270$~MeV$)^3$, the region in which there exists a local inhomogeneous minimum of the thermodynamical potential is extremely narrow{\textemdash}third, top-down phase diagram in Fig.~2 of Ref.~\cite{Carlomagno:2015nsa}. 
In the following we explore the dynamics of the order parameters for values of $T$ and $\mu$ close to the LP and TCP in the equilibrium phase diagram, namely $(T,\mu) = (50~{\rm MeV},190~{\rm MeV})$. 
At this point of the phase diagram, the $\sigma(x)=\sigma\sim250$~MeV and $\pi(x)=0$.

\begin{figure}[htb]
\centering
\includegraphics[width=0.45\textwidth]{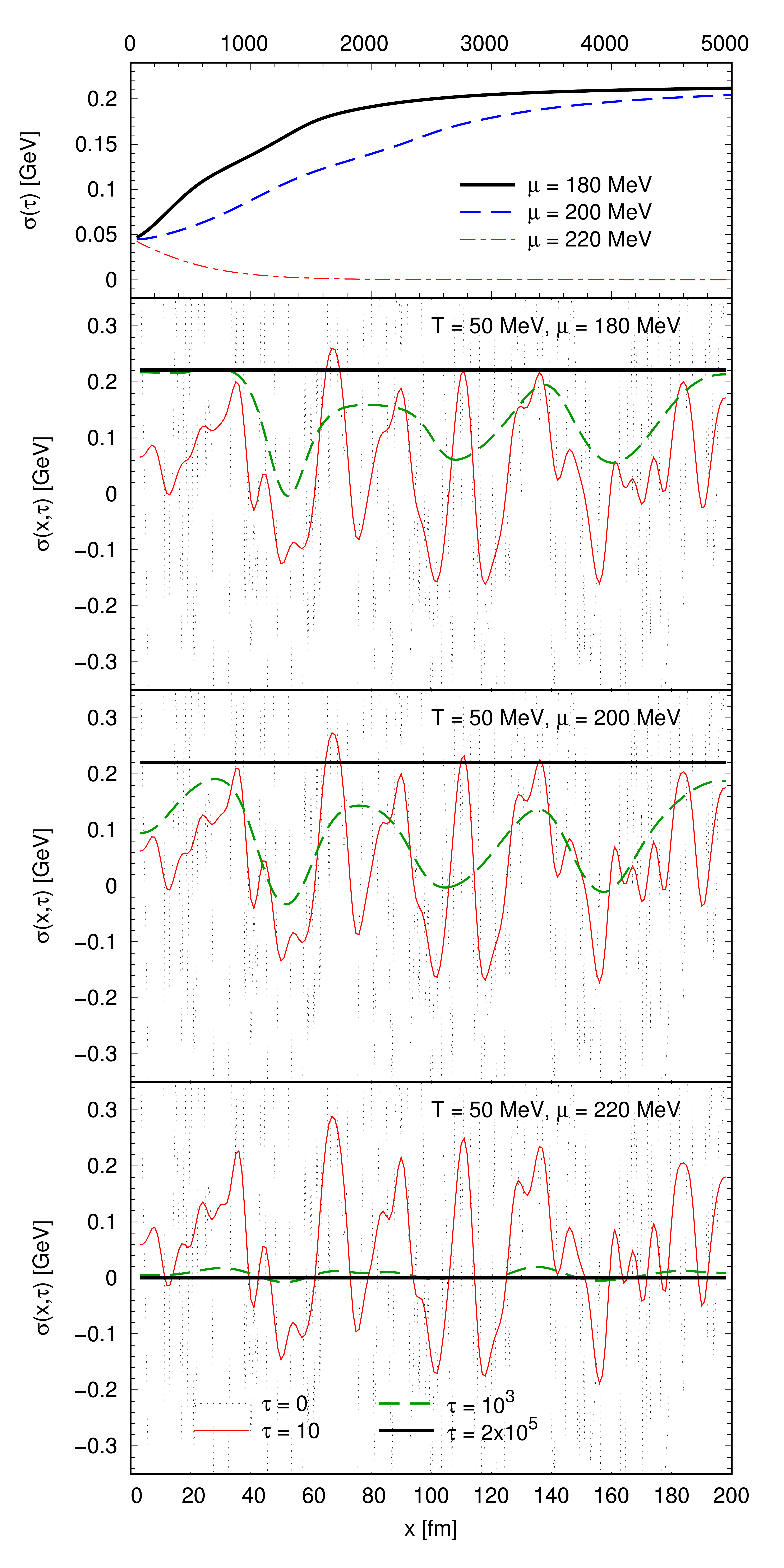}
\caption{\label{fig:ch_rest} Top panel: average of $\sigma(x,\tau)$ over the volume, Eq.~(\ref{av-sig}). Lower panels: snapshots of $\sigma(x,\tau)$ at different values of $\tau$ and ~$\mu$. The initial configuration is a Gaussian profile (light-grey dotted line). }
\end{figure}

Eqs.~(\ref{gllang1}) and (\ref{gllang2}) are solved numerically using a finite-difference method.
We define the dimensionless time coordinate $\tau = t/\Gamma$ and discretize it in time steps $\delta\tau = 10^{-5}$. 
The space coordinate is discretized into a one-dimensional lattice with lattice spacing~$a = 0.1$~fm. 
Since we do not have a microscopic derivation of Eq.~(\ref{tdgle}), the value of $\Gamma$ must be taken from independent sources. 
A good source are calculations using the influence functional in the linear sigma model~\cite{Rischke:1998qy,Nahrgang:2011mv,Nahrgang:2011mg}, where $\Gamma_\sigma$ and $\Gamma_\pi$ as a function of $T$ have been provided. 
For low values of $T$, within the range $0 \leq T \leq 50$~MeV, which is the one of interest here, Ref.~\cite{Rischke:1998qy} finds in the chiral limit $\Gamma_\sigma \simeq 3.75~{\rm fm}$ and $\Gamma_\pi \simeq 0$ (at lowest order the coupling, $\Gamma_\pi = 0$ at all temperatures in the chiral limit). 
The reason for the difference between the coefficients is due to the fact that the process $\sigma \rightarrow 2\pi$ is allowed for all temperatures (including $T =0$), while the reverse process and $\pi \sigma \rightarrow \pi$ are strongly suppressed due to the large mass of the $\sigma$ compared to the one of the $\pi$~\cite{Rischke:1998qy}. 
There is, however, one difficulty here, in that there have been no estimates of $\Gamma$ as a function of $T$ {\em and} $\mu$. In the absence of such estimates, we explore the consequences of using either $\Gamma_\pi = \Gamma_\sigma \neq 0$ and $\Gamma_\sigma \neq 0$ and $\Gamma_\pi = 0$. 

In the top panel of Fig.~\ref{fig:ch_rest} we display the volume average of the scalar field $\sigma(\tau)$:
\be
\phi(\tau) = \frac{1}{N}\sum_{n=1}^N \phi(x_n,\tau) ,
\label{av-sig}
\ee 
where $x_n = n\, a$ and $N$ is the number of lattice sites. 
Results are shown for $T=50$~MeV and three different values of chemical potential: $\mu=180$, $200$ and $220$~MeV. 
Whereas in the remaining panels we plot snapshots of the TDGL time evolution at $\tau=0,\ 10,\ 10^3,\ 2\times10^5$ for the scalar field $\sigma(x,\tau)$. 
The initial profile for $\sigma(x,0)$ and $\pi(x,0)$ was set by imposing a unbiased white Gaussian noise for each position on the lattice, simulating a situation of an out-of-equilibrium state that has been quenched to a low-temperature phase. 
Here, and up to Fig.~\ref{fig:coup_decoup}, we use $\Gamma_\pi = \Gamma_\sigma$. 
We also mention that the pseudoscalar field $\pi(x,\tau)$ (not shown in the figure), after passing through inhomogeneous configurations, evolves to $\pi(x,\tau_{eq})=0$, as it should. 
In addition, it is clear that chiral symmetry is restored for $\mu \simeq 220$~MeV, which is the correct equilibrium state for this value of temperature. 

\begin{figure}[tb]
\includegraphics[width=0.45\textwidth]{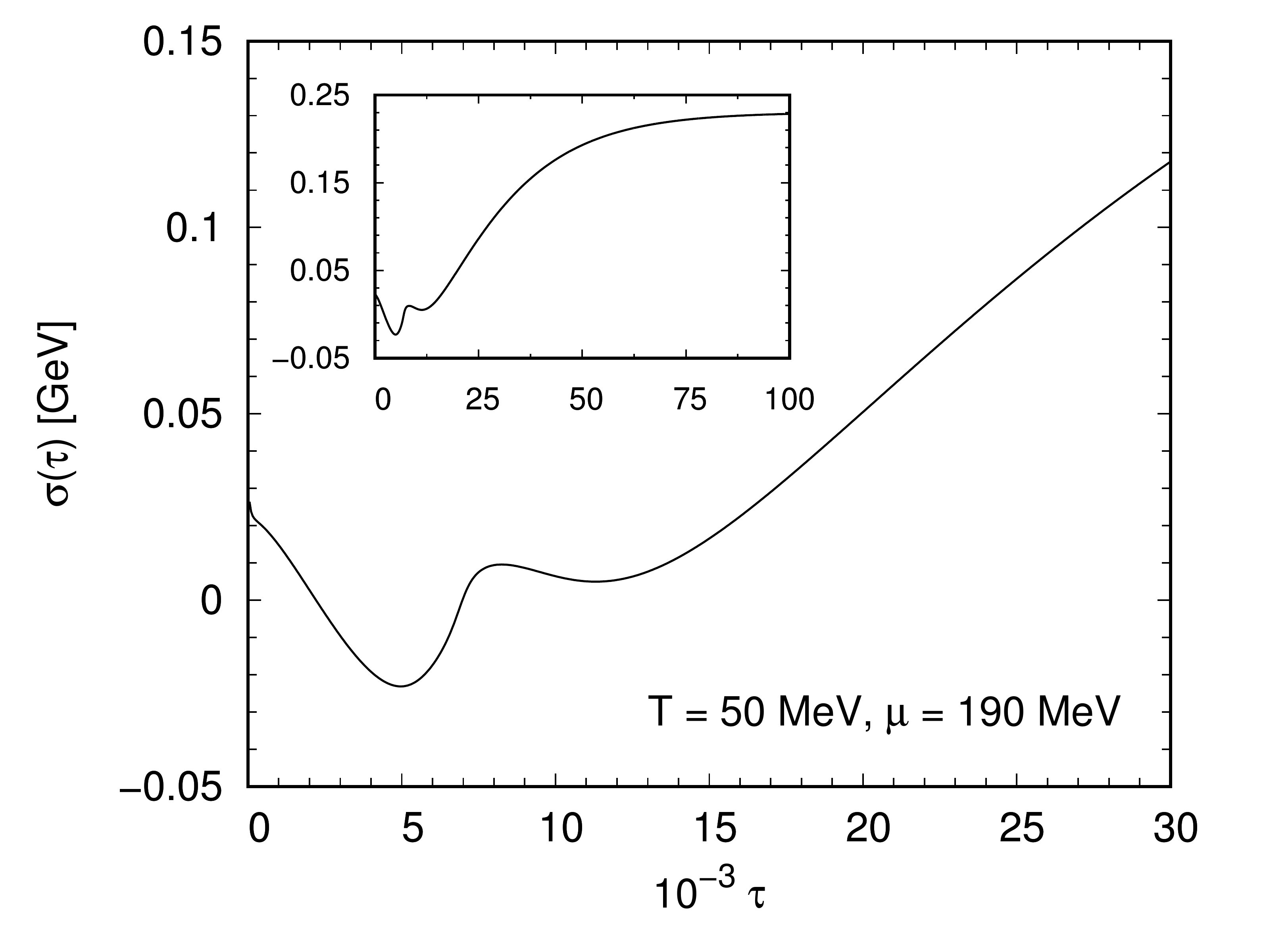} 
\includegraphics[width=0.45\textwidth]{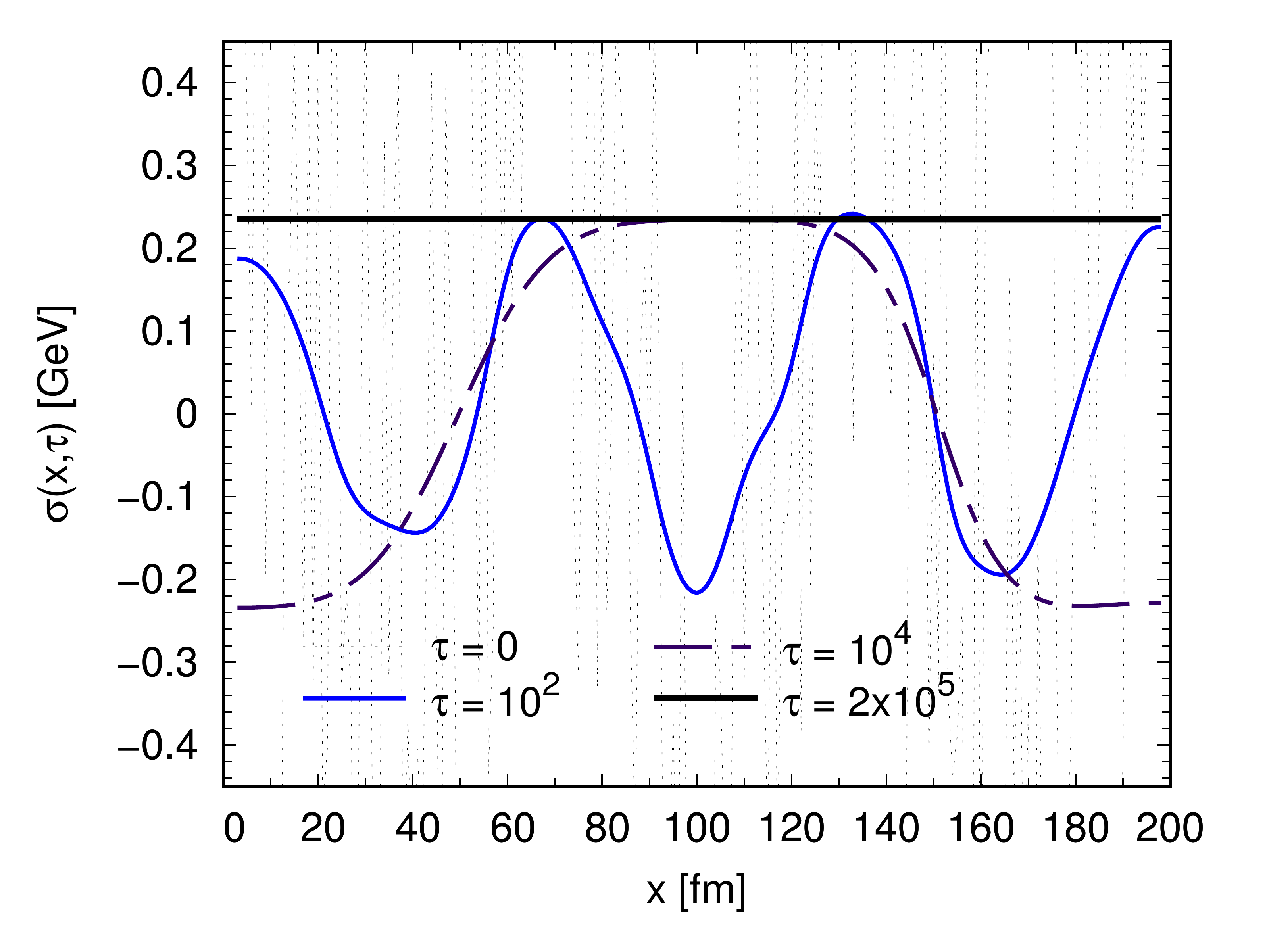}
\caption{\label{fig:inh_dom} Top panel:  average of $\sigma(x,\tau)$ over the volume. Bottom panel: snapshots of $\sigma(x,\tau)$ at different values of $\tau$. The initial configuration is a dual chiral density wave superimposed with Gaussian noise (light-grey dotted line).}
\end{figure}

\begin{figure}[tb]
\centering
\includegraphics[width=0.45\textwidth]{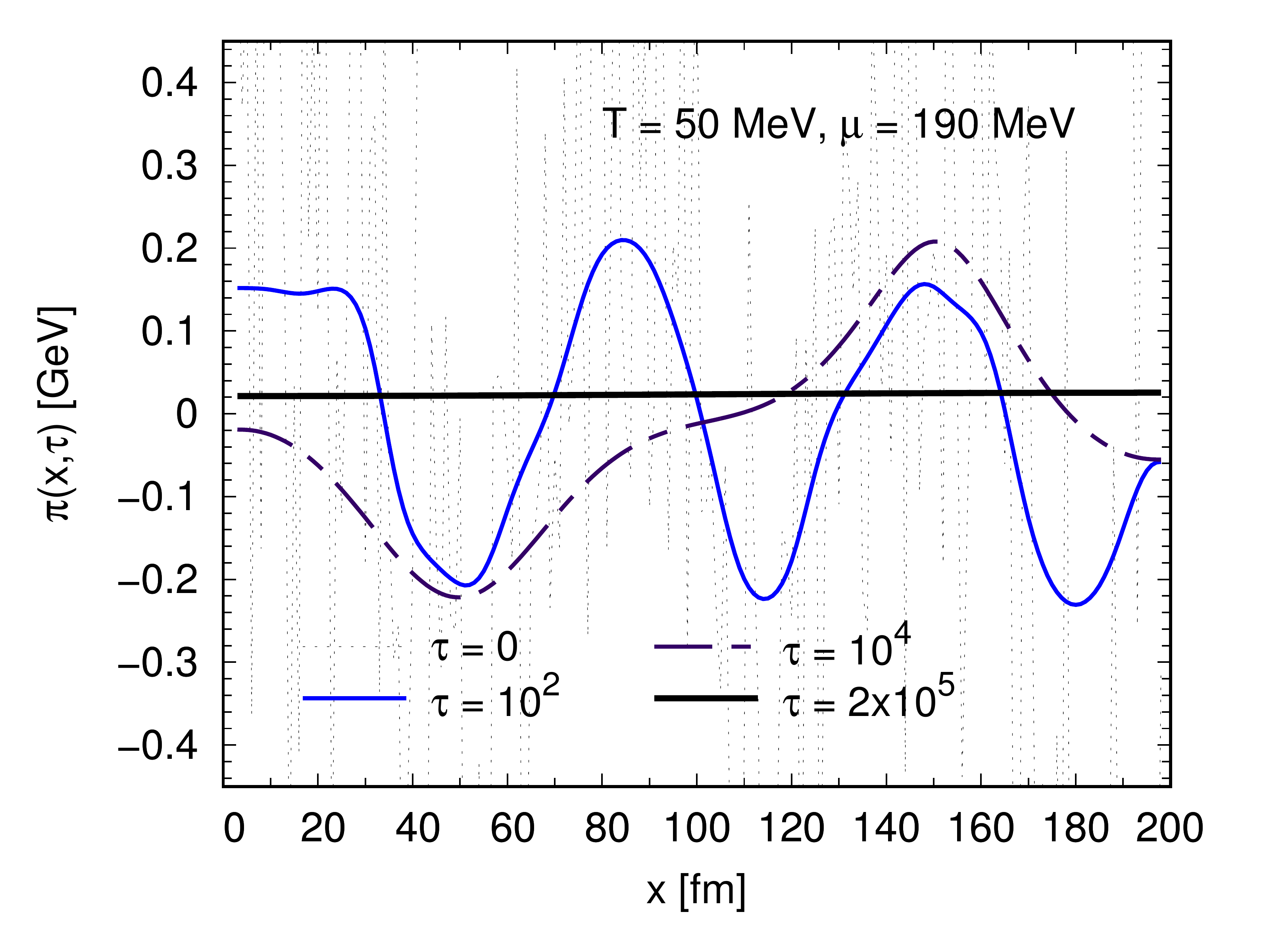}
\caption{\label{fig:pion} TDGL evolution for the pseudoscalar field for a DCDW initial profile superimposed with Gaussian noise (light-grey dotted line).}
\end{figure}

\begin{figure}[t]
\centering
\includegraphics[width=0.45\textwidth]{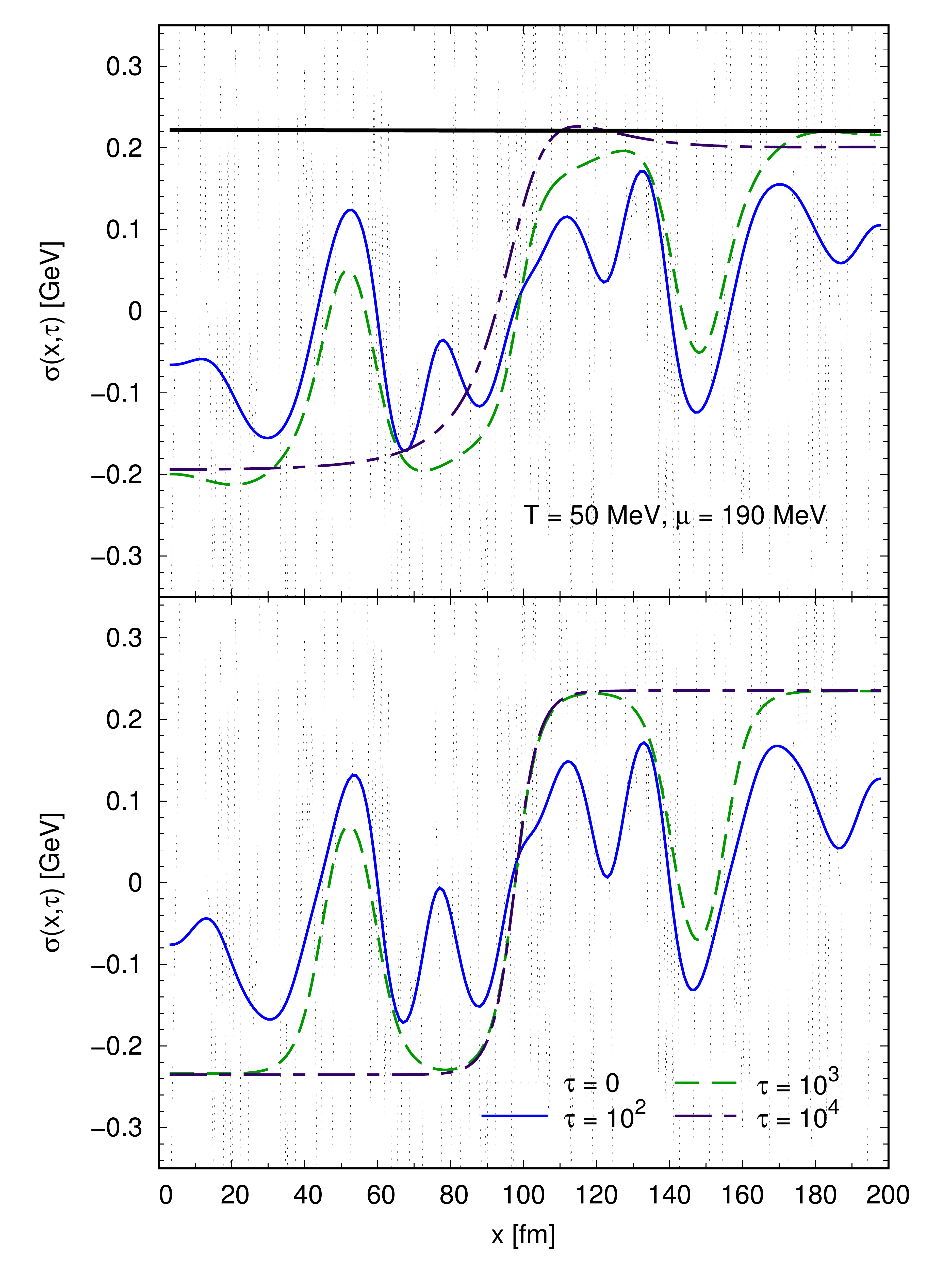}
\caption{\label{fig:coup_decoup} Formation of inhomogeneous configurations for an antisymmetric random initial profile (light-gray dashed lines). When $\Gamma_\pi = 0$, the equilibrium state is an isolated soliton configuration (bottom panel). When the order parameters are coupled, metastable intermediate profiles are formed before reaching a homogeneous configuration at 
equilibrium (top panel). }
\end{figure}

Next, we investigate the effect of the initial configuration on the equilibration time. 
Fig.~\ref{fig:inh_dom} shows results for initial profiles for the scalar and pseudoscalar fields given by dual chiral density waves (DCDW) superimposed with Gaussian noise. 
This kind of modulation is one of the few one-dimensional spatial dependences that can be derived analytically from the GL mean field equations~\cite{Buballa:2014tba}. 
As expected, for the point $(T,\mu) = (50~{\rm MeV}, 190~{\rm MeV})$ of the phase diagram, the equilibrium profile is a homogeneous symmetry-broken configuration. 
One sees that the time to reach the homogeneous equilibrium configuration is longer than in the previous case. 
Here, one also sees an interesting feature of the volume average: for this initial configuration the system stays longer in metastable regions at early times than for the purely random configuration. 
It should be clear that if the homogeneous symmetry-broken solution would be reached without passing through intermediate inhomogeneous phases, the sigma field $\sigma(x,\tau)$ would grow uniformly throughout the volume and, therefore, the volume average $\sigma(\tau)$ would present a monotonic behavior. C
learly, this is not the case. This oscillating behavior of the volume average signals inhomogeneities during the TDGL dynamics, compatible with a DCDW. 
For completeness, we show in Fig.~\ref{fig:pion} the evolution of the pseudoscalar field when the initial condition is a DCDW.
Before equilibrium, this field also goes through inhomogeneous metastable phases and reaches its equilibrium value $\pi(x,t_{eq}) = 0$ for long times. 
It should be noted that the comparable equilibration time for both fields is due to the fact that we are using $\Gamma_\pi = \Gamma_\sigma$ here.

Similar metastable configurations appear in the course of the evolution for different initial conditions. 
The form of the intermediate-state configurations reflect the initial profile.
This is emphasized in Fig.~\ref{fig:coup_decoup}, where we show snapshots of inhomogeneous configurations of $\sigma(x,\tau)$ for an initial antisymmetric random profile. 
The top panel of the figure reveals a hyperbolic tangent profile at long times, before reaching the expected homogeneous configuration (at $\tau \sim 10^5$). 
Next we show the effect of taking $\Gamma_\sigma \neq 0$ and $\Gamma_\pi =0$: the bottom panel of the figure shows that the equilibrium configuration has a hyperbolic tangent shape{\textemdash}increasing the simulation time does not change this shape.
This {\em is not} the expected equilibrium configuration, as discussed previously. 
The point is that in this case of decoupled order parameters, and for this particular initial configuration, the $\sigma$ field is driven into a local minimum of the energy functional. 
The coupling with the $\pi$ field is essential to get the $\sigma$ out of the local minimum. 
Of course, fluctuations, even at such low temperatures, might play a role here and change the picture.

Since the thermalization time depends on the size of the system, it is important to obtain a scaling law of $\tau_{eq}$ with the length of the lattice, $L = N\,a$, where $N$ is the number of lattice points. 
For all the three initial configurations employed along the work, we found that $\tau_{eq}$ grows almost quadratically with $L$.
A good fit to the data from simulations using several values of $L$ is obtained by the formula $\tau_{eq} = A\, L^B$, with $A=3.45(69)$ and $B=2.154(38)$, with a coefficient of determination of $R^2 = 0.993$.  
With such a scaling law, one can make a rough comparison of $\tau_{eq}$ with time scales in typical heavy-ion collision. 
We emphasize that such a comparison is far from rigorous, and it might even not be entirely appropriate given the setting of the study: one-dimensional lattice, no expansion of the system and therefore fixed temperature, etc. 
Nevertheless, for orientation, but keeping this proviso in perspective, if one takes $\Gamma_\sigma = 3.75~{\rm fm}$~\cite{Rischke:1998qy}, well-defined equilibrium configurations are reached after $t_{eq} \sim 10^4~{\rm fm}$, for $L=20$. 
That is, $\tau_{eq}$ is three orders of magnitude larger than, say the decoupling time extracted in Pb-Pb collisions at the LHC~\cite{Foka:2016vta}, for which the kinetic freeze-out volume is $L^3 \sim 5 \times 10^3~{\rm fm}^3$. 
Now, it is important to mention that $(T,\mu) = (50~{\rm MeV}, 190~{\rm MeV})$ corresponds to a baryon density of the order of the nuclear matter density~$\rho_0 = 0.17~{\rm fm}^{-3}$. 
In a heavy-ion collision at FAIR, for example, after a time interval of the order of $15$~fm after the collision, the density of the produced matter will be close to $\rho_0$~\cite{Senger:2017nvf}. 
Therefore, if long-lived chiral inhomogeneities are produced from those high-density regions in such heavy-ion collisions, 
they should leave traces in observables, like e.g. in pion number fluctuations. 
As already mentioned, heavy-ion collisions producing large baryon densities at low temperatures are envisaged at NICA~\cite{Senger:2016wfb} and, more in the future, at J-PARC~\cite{Sako:2017chd}.

%
\section{Summary and conclusions}
\label{finale}

In this work we have analyzed the dynamics of formation of inhomogeneous metastable chiral structures in the symmetry-broken phase of the phase diagram of strongly interacting matter as predicted by a chiral quark model. 
More specifically, we have employed a time-dependent Ginzburg-Landau (TDGL) equation to describe the dynamics of the scalar $\sigma$ and pseudoscalar $\pi$ chiral order parameters near the tricritical point (TCP) and the Lifshitz point (LP) of the equilibrium phase diagram. 
The former denotes the location in the phase diagram where a second order chiral phase transition turns into a first order transition, while the latter determines where one inhomogeneous phase and two homogeneous phases with broken and restored chiral symmetry meet. 
We have used the simplest nonlocal extension of a chiral SU(2) Nambu--Jona-Lasinio model. 
To the best of our knowledge, this work constitutes the first investigation of the dynamics of formation of inhomogeneous chiral phases QCD matter at low temperature and baryon density close to the saturation density of nuclear matter. 

The solutions of the TDGL equations revealed the presence of long-lived metastable configurations of the order parameters in the course of the evolution. 
The time dynamic was studied in a region of the phase diagram where the equilibrium configurations of the $\sigma$ and $\pi$ order parameters have no spatial modulations. 
Initially, we verified our approach in regard to chiral restoration at low temperature as function of the chemical potential $\mu$ in the vicinity of the critical points. 
We verified that the TDGL equation leads an equilibrium solution for which the symmetry is restored for values of $\mu$ in good agreement with those obtained in Ref.~\cite{Carlomagno:2015nsa}. 
We also found that when $\pi$ is set equal to zero during the entire evolution, the $\sigma$ field is not driven to the correct equilibrium configuration; it is driven into a local minimum of the energy functional. 
The coupling with the $\pi$ field is essential to drive $\sigma$ to the correct equilibrium configuration.  

To finalize, the main conclusion of the paper is that inhomogeneous configurations of the chiral order parameters produced in the course of the evolution of matter can be long-lived, with lifetimes much larger than the typical lifetimes in a heavy-ion collision. 
This means that, even if at equilibrium the chiral parameters have no spatial modulation, the system decouples before reaching such a state and can leave traces of the inhomogeneities in observables. 
The investigation of such possible traces in observable is left for a future study, when several extensions of the present study will be made. 
First of all, it is underway the extension of the study to three-dimensions and inclusion of expansion of the system~\cite{Carlomagno:2018lcc}. 
Finally, the derivation of a TDGL equation from the nonlocal NJL model used in the present work would be essential to avoid uncertainties regarding the kinetic coefficients.

%
\section*{Acknowledgements}

This work has been partially funded by CONICET under Grant No.\ PIP 449 (J.P.C.), the National University of La Plata, Project No.\ X718 (J.P.C.), Conselho Nacional de Desenvolvimento Cient\'{\i}fico e e Tecnol\'ogico - CNPq, Grant. No. 305894/2009-9 (G.K.), and Fundac\c{c}\~ao de Amparo \`a Pesquisa do Estado de S\~ao Paulo - FAPESP, Grant Nos. 2013/01907-0 (G.K.) and 2016/01343-7 (J.P.C.).




\begin{thebibliography}{99}
%
\bibitem{Fukushima:2010bq} 
  K.~Fukushima and T.~Hatsuda,
  Rept.\ Prog.\ Phys.\  {\bf 74}, 014001 (2011)
  doi:10.1088/0034-4885/74/1/014001
  [arXiv:1005.4814 [hep-ph]].
%
\bibitem{Holt:2014hma} 
  J.~W.~Holt, M.~Rho and W.~Weise,
  Phys.\ Rept.\  {\bf 621}, 2 (2016)
  doi:10.1016/j.physrep.2015.10.011
  [arXiv:1411.6681 [nucl-th]].
%
\bibitem{Baym:2017whm} 
  G.~Baym, T.~Hatsuda, T.~Kojo, P.~D.~Powell, Y.~Song and T.~Takatsuka,
  arXiv:1707.04966 [astro-ph.HE].
%
\bibitem{Blaschke:2018mqw} 
  D.~Blaschke and N.~Chamel,
  arXiv:1803.01836 [nucl-th].
%
\bibitem{Splittorff:2007ck} 
  K.~Splittorff and J.~J.~M.~Verbaarschot,
  Phys.\ Rev.\ D {\bf 75}, 116003 (2007)
  doi:10.1103/PhysRevD.75.116003
  [hep-lat/0702011 [HEP-LAT]].
%
\bibitem{Aarts:2015tyj} 
  G.~Aarts,
  J.\ Phys.\ Conf.\ Ser.\  {\bf 706}, no. 2, 022004 (2016)
  doi:10.1088/1742-6596/706/2/022004
  [arXiv:1512.05145 [hep-lat]].
%
\bibitem{Aoki:2006we} 
  Y.~Aoki, G.~Endrodi, Z.~Fodor, S.~D.~Katz and K.~K.~Szabo,
  Nature {\bf 443}, 675 (2006)
  doi:10.1038/nature05120
  [hep-lat/0611014].
%
\bibitem{Cheng:2007jq} 
  M.~Cheng {\it et al.},
  Phys.\ Rev.\ D {\bf 77}, 014511 (2008)
  doi:10.1103/PhysRevD.77.014511
  [arXiv:0710.0354 [hep-lat]].
%
\bibitem{Bhattacharya:2014ara} 
  T.~Bhattacharya {\it et al.},
  Phys.\ Rev.\ Lett.\  {\bf 113}, no. 8, 082001 (2014)
  doi:10.1103/PhysRevLett.113.082001
  [arXiv:1402.5175 [hep-lat]].
%
\bibitem{Nambu:1960xd} 
  Y.~Nambu,
  Phys.\ Rev.\ Lett.\  {\bf 4}, 380 (1960).
  doi:10.1103/PhysRevLett.4.380
%
\bibitem{Horn:2016rip} 
  T.~Horn and C.~D.~Roberts,
  J.\ Phys.\ G {\bf 43}, no. 7, 073001 (2016)
  doi:10.1088/0954-3899/43/7/073001
  [arXiv:1602.04016 [nucl-th]].
%
\bibitem{Nakano:2004cd} 
  E.~Nakano and T.~Tatsumi,
  Phys.\ Rev.\ D {\bf 71}, 114006 (2005)
  doi:10.1103/PhysRevD.71.114006
  [hep-ph/0411350].
%
\bibitem{Basar:2008im} 
  G.~Basar and G.~V.~Dunne,
  Phys.\ Rev.\ Lett.\  {\bf 100}, 200404 (2008)
  doi:10.1103/PhysRevLett.100.200404
  [arXiv:0803.1501 [hep-th]].
%
\bibitem{Basar:2008ki} 
  G.~Basar and G.~V.~Dunne,
  Phys.\ Rev.\ D {\bf 78}, 065022 (2008)
  doi:10.1103/PhysRevD.78.065022
  [arXiv:0806.2659 [hep-th]].
%
\bibitem{Nickel:2009ke} 
  D.~Nickel,
  Phys.\ Rev.\ Lett.\  {\bf 103}, 072301 (2009)
  doi:10.1103/PhysRevLett.103.072301
  [arXiv:0902.1778 [hep-ph]].
%
\bibitem{Basar:2009fg} 
  G.~Basar, G.~V.~Dunne and M.~Thies,
  Phys.\ Rev.\ D {\bf 79}, 105012 (2009)
  doi:10.1103/PhysRevD.79.105012
  [arXiv:0903.1868 [hep-th]].
%
\bibitem{Nickel:2009wj} 
  D.~Nickel,
  Phys.\ Rev.\ D {\bf 80}, 074025 (2009)
  doi:10.1103/PhysRevD.80.074025
  [arXiv:0906.5295 [hep-ph]].
%
\bibitem{Buballa:2014tba} 
  M.~Buballa and S.~Carignano,
  Prog.\ Part.\ Nucl.\ Phys.\  {\bf 81}, 39 (2015)
  doi:10.1016/j.ppnp.2014.11.001
  [arXiv:1406.1367 [hep-ph]].
%
\bibitem{Carignano:2014jla} 
  S.~Carignano, M.~Buballa and B.~J.~Schaefer,
  Phys.\ Rev.\ D {\bf 90}, no. 1, 014033 (2014)
  doi:10.1103/PhysRevD.90.014033
  [arXiv:1404.0057 [hep-ph]].
%
\bibitem{Heinz:2013hza}
  A.~Heinz, F.~Giacosa and D.~H.~Rischke,
  Nucl.\ Phys.\ A {\bf 933} (2015) 34
  doi:10.1016/j.nuclphysa.2014.09.027
  [arXiv:1312.3244 [nucl-th]].
%
\bibitem{Tatsumi:2014cea} 
  T.~Tatsumi and T.~Muto,
  Phys.\ Rev.\ D {\bf 89}, 103005 (2014).
%
\bibitem{Carignano:2015kda} 
  S.~Carignano, E.~J.~Ferrer, V.~de la Incera and L.~Paulucci,
  Phys.\ Rev.\ D {\bf 92}, 105018 (2015).
%
\bibitem{Buballa:2015awa} 
  M.~Buballa and S.~Carignano,
  Eur.\ Phys.\ J.\ A {\bf 52}, 57 (2016).
%
\bibitem{Kumar:2013cqa}
  L.~Kumar,
  Mod.\ Phys.\ Lett.\ A {\bf 28} (2013) 1330033
  doi:10.1142/S0217732313300334
  [arXiv:1311.3426 [nucl-ex]].
%
\bibitem{Senger:2017nvf} 
  P.~Senger,
  J.\ Phys.\ Conf.\ Ser.\  {\bf 798}, no. 1, 012062 (2017).
  doi:10.1088/1742-6596/798/1/012062
%
\bibitem{Senger:2016wfb} 
  P.~Senger,
  Eur.\ Phys.\ J.\ A {\bf 52}, no. 8, 217 (2016).
  doi:10.1140/epja/i2016-16217-8
%
\bibitem{Sako:2017chd}
  H.~Sako,
  PoS CPOD {\bf 2017} (2018) 028.
%
\bibitem{Yoshiike:2017kbx} 
  R.~Yoshiike, T.~G.~Lee and T.~Tatsumi,
  Phys.\ Rev.\ D {\bf 95}, no. 7, 074010 (2017)
  doi:10.1103/PhysRevD.95.074010
  [arXiv:1702.01511 [hep-ph]].
%
\bibitem{Onuki:2002} 
  A.~Onuki,
  Phase Transition Dynamics,
  (Cambridge University Press, 2002)
%
\bibitem{Bray:1994zz} 
  A.~J.~Bray,
  Adv.\ Phys.\  {\bf 43}, 357 (1994).
  doi:10.1080/00018739400101505
%
\bibitem{Csernai:1995zn} 
  L.~P.~Csernai and I.~N.~Mishustin,
  Phys.\ Rev.\ Lett.\  {\bf 74}, 5005 (1995).
  doi:10.1103/PhysRevLett.74.5005
%
\bibitem{Biro:1997va} 
  T.~S.~Biro and C.~Greiner,
  Phys.\ Rev.\ Lett.\  {\bf 79}, 3138 (1997)
  doi:10.1103/PhysRevLett.79.3138
  [hep-ph/9704250].
%
\bibitem{Mishustin:1998yc} 
  I.~N.~Mishustin and O.~Scavenius,
  Phys.\ Rev.\ Lett.\  {\bf 83}, 3134 (1999)
  doi:10.1103/PhysRevLett.83.3134
  [hep-ph/9804338].
%
\bibitem{Scavenius:1999zc} 
  O.~Scavenius and A.~Dumitru,
  Phys.\ Rev.\ Lett.\  {\bf 83}, 4697 (1999)
  doi:10.1103/PhysRevLett.83.4697
  [hep-ph/9905572].
%
\bibitem{Miller:2000pd}
  T.~R.~Miller and M.~C.~Ogilvie,
  Phys.\ Lett.\ B {\bf 488} (2000) 313
  doi:10.1016/S0370-2693(00)00922-9
  [hep-lat/0004004].
%
\bibitem{Bower:2001fq}
  D. Bower and S. Gavin,
  Phys. Rev. C {\bf 64}, 051902 (2001).
%
\bibitem{Scavenius:2001pa} 
  O.~Scavenius, A.~Dumitru and A.~D.~Jackson,
  Phys.\ Rev.\ Lett.\  {\bf 87}, 182302 (2001)
  doi:10.1103/PhysRevLett.87.182302
  [hep-ph/0103219].
%
\bibitem{Paech:2003fe} 
  K.~Paech, H.~Stoecker and A.~Dumitru,
  Phys.\ Rev.\ C {\bf 68}, 044907 (2003)
  doi:10.1103/PhysRevC.68.044907
  [nucl-th/0302013].
%
\bibitem{Aziz:2004qu} 
  M.~A.~Aziz and S.~Gavin,
  Phys.\ Rev.\ C {\bf 70}, 034905 (2004)
  doi:10.1103/PhysRevC.70.034905
  [nucl-th/0404058].
%
\bibitem{Fraga:2004hp} 
  E.~S.~Fraga and G.~Krein,
  Phys.\ Lett.\ B {\bf 614}, 181 (2005).
%
\bibitem{Paech:2005cx} 
  K.~Paech and A.~Dumitru,
  Phys.\ Lett.\ B {\bf 623}, 200 (2005)
  doi:10.1016/j.physletb.2005.08.006
  [nucl-th/0504003].
%
\bibitem{Fraga:2006cr} 
  E.~S.~Fraga, T.~Kodama, G.~Krein, A.~J.~Mizher and L.~F.~Palhares,
  Nucl.\ Phys.\ A {\bf 785}, 138 (2007).
%
\bibitem{Koide:2006vf} 
  T.~Koide, G.~Krein and R.~O.~Ramos,
  Phys.\ Lett.\ B {\bf 636}, 96 (2006)
  doi:10.1016/j.physletb.2006.03.035
  [hep-ph/0601256].
%
\bibitem{Fraga:2007gg} 
  E.~S.~Fraga, G.~Krein and A.~J.~Mizher,
  Phys.\ Rev.\ D {\bf 76}, 034501 (2007).
%
\bibitem{CassolSeewald:2007ru} 
  N.~C.~Cassol-Seewald, R.~L.~S.~Farias, E.~S.~Fraga, G.~Krein and R.~O.~Ramos,
  Physica A {\bf 391}, 4088 (2012).
%
\bibitem{Randrup:2010ax} 
  J.~Randrup,
  Phys.\ Rev.\ C {\bf 82}, 034902 (2010)
  doi:10.1103/PhysRevC.82.034902
  [arXiv:1007.1448 [nucl-th]].
%
\bibitem{Singh:2011rz} 
  A.~Singh, S.~Puri and H.~Mishra,
  Nucl.\ Phys.\ A {\bf 864}, 176 (2011)
  doi:10.1016/j.nuclphysa.2011.06.023
  [arXiv:1101.0500 [hep-ph]].
%
\bibitem{Kapusta:2012zb}
  J.~I.~Kapusta and J.~M.~Torres-Rincon,
  Phys.\ Rev.\ C {\bf 86} (2012) 054911
  doi:10.1103/PhysRevC.86.054911
  [arXiv:1209.0675 [nucl-th]].
%
\bibitem{Singh:2012nq} 
  A.~Singh, S.~Puri and H.~Mishra,
  Nucl.\ Phys.\ A {\bf 908}, 12 (2013)
  doi:10.1016/j.nuclphysa.2013.03.016
  [arXiv:1209.6137 [hep-ph]].
%
\bibitem{Singh:2013pxa} 
  A.~Singh, S.~Puri and H.~Mishra,
  EPL {\bf 102}, no. 5, 52001 (2013)
  doi:10.1209/0295-5075/102/52001
  [arXiv:1306.5047 [hep-ph]].
%
\bibitem{Carlomagno:2014hoa} 
  J.~P.~Carlomagno, D.~Gomez Dumm and N.~N.~Scoccola,
  Phys.\ Lett.\ B {\bf 745}, 1 (2015)
  doi:10.1016/j.physletb.2015.04.023
  [arXiv:1411.0909 [hep-ph]].
%
\bibitem{Carlomagno:2015nsa} 
  J.~P.~Carlomagno, D.~G\'omez Dumm and N.~N.~Scoccola,
  Phys.\ Rev.\ D {\bf 92}, 056007 (2015).
%
\bibitem{Calzetta:2008iqa}
  E.~A.~Calzetta and B.~L.~B.~Hu,
  Nonequilibrium Quantum Field Theory (Cambridge University Press, Cambridge, 2008)
%
\bibitem{Gleiser:1993ea} 
  M.~Gleiser and R.~O.~Ramos,
  Phys.\ Rev.\ D {\bf 50}, 2441 (1994).
%
\bibitem{Boyanovsky:1994me} 
  D.~Boyanovsky, H.~J.~de Vega, R.~Holman, D.~S.~Lee and A.~Singh,
  Phys.\ Rev.\ D {\bf 51}, 4419 (1995)
  doi:10.1103/PhysRevD.51.4419
  [hep-ph/9408214].
%
\bibitem{Greiner:1996dx} 
  C.~Greiner and B.~Muller,
  Phys.\ Rev.\ D {\bf 55}, 1026 (1997)
  doi:10.1103/PhysRevD.55.1026
  [hep-th/9605048].
%
\bibitem{Greiner:1998vd} 
  C.~Greiner and S.~Leupold,
  Annals Phys.\  {\bf 270}, 328 (1998)
  doi:10.1006/aphy.1998.5849
  [hep-ph/9802312].
%
\bibitem{Rischke:1998qy} 
  D.~H.~Rischke,
  Phys.\ Rev.\ C {\bf 58}, 2331 (1998).
%
\bibitem{Nahrgang:2011mv} 
  M.~Nahrgang, S.~Leupold and M.~Bleicher,
  Phys.\ Lett.\ B {\bf 711}, 109 (2012).
%
\bibitem{Nahrgang:2011mg} 
  M.~Nahrgang, S.~Leupold, C.~Herold and M.~Bleicher,
  Phys.\ Rev.\ C {\bf 84}, 024912 (2011).
%
\bibitem{Gautier:2012vh} 
  F.~Gautier and J.~Serreau,
  Phys.\ Rev.\ D {\bf 86}, 125002 (2012)
  doi:10.1103/PhysRevD.86.125002
  [arXiv:1209.1827 [hep-th]].
%
\bibitem{Fulde:1964zz}
  P.~Fulde and R.~A.~Ferrell,
  Phys.\ Rev.\  {\bf 135} (1964) A550.
  doi:10.1103/PhysRev.135.A550
%
\bibitem{Alford:2000ze}
  M.~G.~Alford, J.~A.~Bowers and K.~Rajagopal,
  Phys.\ Rev.\ D {\bf 63} (2001) 074016
  doi:10.1103/PhysRevD.63.074016
  [hep-ph/0008208].
%
\bibitem{Abuki:2011pf} 
  H.~Abuki, D.~Ishibashi and K.~Suzuki,
  Phys.\ Rev.\ D {\bf 85}, 074002 (2012)
  doi:10.1103/PhysRevD.85.074002
  [arXiv:1109.1615 [hep-ph]].
%
\bibitem{Carignano:2017meb}
  S.~Carignano, F.~Anzuini, O.~Benhar and M.~Mannarelli,
  arXiv:1711.08607 [hep-ph].
%
\bibitem{Bowler:1994ir} 
  R.~D.~Bowler and M.~C.~Birse,
  Nucl.\ Phys.\ A {\bf 582}, 655 (1995)
  doi:10.1016/0375-9474(94)00481-2
  [hep-ph/9407336].
%
\bibitem{Schmidt:1994di} 
  S.~M.~Schmidt, D.~Blaschke and Y.~L.~Kalinovsky,
  Phys.\ Rev.\ C {\bf 50}, 435 (1994).
  doi:10.1103/PhysRevC.50.435
%
\bibitem{Golli:1998rf} 
  B.~Golli, W.~Broniowski and G.~Ripka,
  Phys.\ Lett.\ B {\bf 437}, 24 (1998)
  doi:10.1016/S0370-2693(98)00942-3
  [hep-ph/9807261].

\bibitem{GomezDumm:2001fz} 
  D.~Gomez Dumm and N.~N.~Scoccola,
  Phys.\ Rev.\ D {\bf 65}, 074021 (2002)
  doi:10.1103/PhysRevD.65.074021
  [hep-ph/0107251].
%
\bibitem{Scarpettini:2003fj} 
  A.~Scarpettini, D.~Gomez Dumm and N.~N.~Scoccola,
  Phys.\ Rev.\ D {\bf 69}, 114018 (2004)
  doi:10.1103/PhysRevD.69.114018
  [hep-ph/0311030].
%
\bibitem{GomezDumm:2004sr} 
  D.~Gomez Dumm and N.~N.~Scoccola,
  Phys.\ Rev.\ C {\bf 72}, 014909 (2005)
  doi:10.1103/PhysRevC.72.014909
  [hep-ph/0410262].
%
\bibitem{GomezDumm:2006vz} 
  D.~Gomez Dumm, A.~G.~Grunfeld and N.~N.~Scoccola,
  Phys.\ Rev.\ D {\bf 74}, 054026 (2006)
  doi:10.1103/PhysRevD.74.054026
  [hep-ph/0607023].
%
\bibitem{Aoki:2013ldr} 
  S.~Aoki {\it et al.},
  Eur.\ Phys.\ J.\ C {\bf 74}, 2890 (2014)
  doi:10.1140/epjc/s10052-014-2890-7
  [arXiv:1310.8555 [hep-lat]].
%
\bibitem{Foka:2016vta} 
  P.~Foka and M.~A.~Janik,
  Rev.\ Phys.\  {\bf 1}, 154 (2016)
  doi:10.1016/j.revip.2016.11.002
  [arXiv:1702.07233 [hep-ex]].
%
\bibitem{Carlomagno:2018lcc} 
  J.~P.~Carlomagno, G.~Krein, D.~Kroff and T.~Peixoto,
  EPJ Web Conf.\  {\bf 172}, 03005 (2018).
  doi:10.1051/epjconf/201817203005

\end{thebibliography}
\end{document}